\documentclass[fleqn,usenatbib]{mnras}
\usepackage{newtxtext,newtxmath}
\usepackage[T1]{fontenc}
\usepackage{units}
\usepackage{lineno}
\usepackage{color}
\usepackage{epsfig}
\usepackage{float}
\usepackage{bm}
\DeclareRobustCommand{\VAN}[3]{#2}
\let\VANthebibliography\thebibliography
\def\thebibliography{\DeclareRobustCommand{\VAN}[3]{##3}\VANthebibliography}
\usepackage{graphicx}
\usepackage{amsmath}

\usepackage{amssymb}
\usepackage{amsfonts} 
\usepackage{caption}
\usepackage{subfigure}

\title[Periodic Activities of FRB repeaters from Magnetars]{Periodic Activities of Fast Radio Burst Repeaters from Precessing Magnetars with Evolving Obliquity} 

\author[Feng et al.]{
Xin-Ming Feng$^{1}$,
Yuan-Pei Yang$^{1,2}$\thanks{E-mail: ypyang@ynu.edu.cn (YPY)}
and Qiao-Chu Li$^{3,4}$
\\
$^{1}$South-Western Institute for Astronomy Research, Yunnan University, Kunming 650504, China\\
$^{2}$Purple Mountain Observatory, Chinese Academy of Sciences, Nanjing 210023, China\\
$^{3}$School of Astronomy and Space Science, Nanjing University, Nanjing 210023, China\\
$^{4}$Key Laboratory of Modern Astronomy and Astrophysics (Nanjing University), Ministry of Education, Nanjing 210023, China\\
}

\date{Accepted XXX. Received YYY; in original form ZZZ}

\pubyear{2024}

\begin{document}
\label{firstpage}
\pagerange{\pageref{firstpage}--\pageref{lastpage}} 
\maketitle

\begin{abstract}

Fast radio bursts (FRBs) are cosmological radio transients with millisecond durations and extremely high brightness temperatures. One FRB repeater, FRB 180916.J0158+65 (FRB 180916B), was confirmed to appear 16.35-day periodic activities with 5-day activity window. Another FRB repeater, FRB 121102, and two soft gamma-ray repeaters (SGRs), SGR 1935+2154 and SGR 1806-20, also show possible periodic activities. These periodicities might originate from the precession process of young magnetars due to the anisotropic pressure from the inner magnetic fields as proposed in the literature. In this work, we analyze a self-consistent model for the rotation evolution of magnetars and obtain the evolutions of magnetar precession and obliquity. We find that if the FRB repeaters and the SGRs with (possible) periodic activities originate from the magnetar precession, their ages would be constrained to be hundreds to tens of thousands of years, which is consistent with the typical ages of magnetars. Assuming that the FRB emission is beaming in the magnetosphere as proposed in the literature, we calculate the evolution of the observable probability and the duty cycle of the active window period. We find that for a given magnetar the observable probability increases with the magnetar age in the early stage and decreases with the magnetar age in the later stage, meanwhile, there are one or two active windows in one precession period if the emission is not perfectly axisymmetric with respect to the deformation axis of a magnetar, which could be tested by the future observation for repeating FRB sources. 

\end{abstract}


\section{Introduction}\label{sec:intro}

Fast radio bursts (FRBs) are radio transients with millisecond durations and extremely high brightness temperatures. Since the ``Lorimer Burst'' (FRB 010724) was discovered in 2007 \citep{Lorimer2007}, hundreds of FRBs have been published so far, and dozens of them appear repeating behaviors \citep[e.g.,][]{Thornton2013,Spitler2016,Chatterjee2017,Bannister2019,Ravi2019,Prochaska2019,Marcote2020}.
FRBs are found to be almost isotropic distributed in all sky, and the dispersion measures (DMs) of most FRBs exceed the DM contributions from the Milky Way, suggesting a cosmological origin \citep[e.g.,][]{Thornton2013}. Together with the short durations, their brightness temperatures are required to be extremely high, $T_{\rm B}\gtrsim 10^{35}{\rm K}$. Therefore, the radiation mechanism of FRBs must be coherent \citep[e.g.,][]{ZhangBing2020}. Possible coherent radiation mechanisms applying for FRBs include curvature radiation by charged bunches \citep[e.g.,][]{Kumar2017,YangYuanPei2018,Yang2023,LuWenBin2020,YangYuanPei2020,Cooper2021}, synchrotron maser \citep[e.g.,][]{Waxman2017,Metzger2019,Beloborodov2020,Lyubarsky2020}, and plasma radiation \citep[e.g.,][]{Philippov2020,Yang2021,Mahlmann2022}.
On the other hand, the physical origin of FRBs is still not well understood. 
Based on the distance of the emission region from the FRB engine, e.g., a neutron star, there are two broad-classes models to explain their physical origin: 1. FRBs are generated in the magnetosphere of a neutron star \citep{Kumar2017,Katz2018,Lu2018,YangYuanPei2018,Yang2021,Wadiasingh2019,Kumar2020,LuWenBin2020,WangJieShuang2020,WangWY2020,Lyutikov2020,Wadiasingh2020,Zhang2017,Dai2020a,Geng2020,Ioka2020b}. 2. FRBs are produced by the synchrotron maser mechanism in the shock of an outflow far away from the neutron star \citep{Lyubarsky2014,Waxman2017,Metzger2019,Beloborodov2020,Marcote2020,Yu2021,Wu2020}.

The recent CHIME/FRB catalog represented the first large FRB sample, including more than ten FRB repeaters and a few hundreds of one-off FRBs \citep{TheCHIMEFRBCollaboration2021}. It is interesting that the bursts from FRB repeaters have relatively larger pulse duration, narrower bandwidth, and lower brightness temperature compared with the one-off FRBs \citep{Pleunis2021,Luo2022,Zhu-Ge2022}.
Some FRB repeaters appear (possible) periodic activities: 1) FRB 180916.J0158+65 (FRB 180916B) shows a 16.35-day periodic activity with a 5-day activity window \citep{ChimeFrbCollaboration2020,Pastor-Marazuela2021,Bethapudi2022}; 2) FRB 121102 shows a possibly longer period of 161 day with an active window of about 100 days \citep{Rajwade2020,Cruces2021}. Besides, some Galactic magnetars also appear possible periodic activities in SGRs. For example, SGR 1806-20 and SGR 1935+2154 were found to have possible periodic activities of 398.2 days \citep{ZhangGQ2021} and 127 day\footnote{In one earlier work \citep{Zou2021}, a possible 238-day activity period of SGR 1935+2154 was reported using the data of Fermi/GBM from 2014 July to 2021 October.} \citep{Xie2022}, respectively.  

Some astrophysical scenarios are involved in explaining the periodic activities of FRB repeaters (see the review of \citet{ZhangBing2020}, and some detailed comments on different scenarios are also discussed in \citet{Katz2021} and \citet{Wei2021}): 1) an FRB repeater is in a binary star systems, and the activity period corresponds to the orbital period of the binary system \citep{Dai2020b,Ioka2020,LiQiaochu2021}. Such a scenario predicts that some observation properties, e.g., dispersion measure and rotation measure, of the FRB repeater might be periodic if the turbulence in the environment is not significant \citep{WangFayin2022,Yang2022,Zhao2022}. 2) the periodic activity of an FRB repeater is due to the precession of an isolated deformed neutron star, meanwhile, the FRB emission is beaming in the magnetosphere of the neutron star \citep{Levin2020,Zanazzi2020,Sridhar2021,LiDongZi2020,Cordes2021}. 
3) the periodic activity is produced by the forced precession contributed by orbital motion \citep{YangHuan2020} or fallback disk \citep{Tong2020}.
4) periodicity corresponds to the slow rotation of an isolated neutron star, like a radio pulsar \citep{Beniamini2020,Xu2021}. 
5) the periodic activities originate from the thermal-viscous instability of the accretion disk around the neutron star in a binary system \citep{Geng2021}. 
6) the periodicity originates from the precession of the jet or accretion disc of a compact object (neutron star or black hole) \citep{Sridhar2022,Katz2022}.
It is worth noting that although FRB 180916B exhibits a 16.35-day periodicity in its burst activity, its RM appears unrelated to the periodic activity and evolves during a long term much larger than the activity period \citep{Mckinven2022}. Thus, the activity period seems more likely to originate from the intrinsic process of the FRB engine, e.g., the precession or rotation of a compact object.

In this work, we consider that the observed periodic activities of FRB repeaters and SGRs are due to the magnetar precession, and constrain the magnetars' ages under the picture of the self-consistent dynamic evolution of the magnetar precession. We also discuss the observable probability and the activity window due to the obliquity evolution.
The paper is organized as follows. We calculate the evolution of the magnetar rotation and precession in Section \ref{sec2}, and constrain the ages of some special sources, including FRB repeaters and SGRs in Section \ref{sec3}. We discuss the observable probability and the activity window in Section \ref{sec4}. The results are discussed and summarized in Section \ref{sec5}. 

\section{Rotation and precession evolution of magnetars}\label{sec2} 

\begin{figure}
\centering
\includegraphics[width = 8cm]{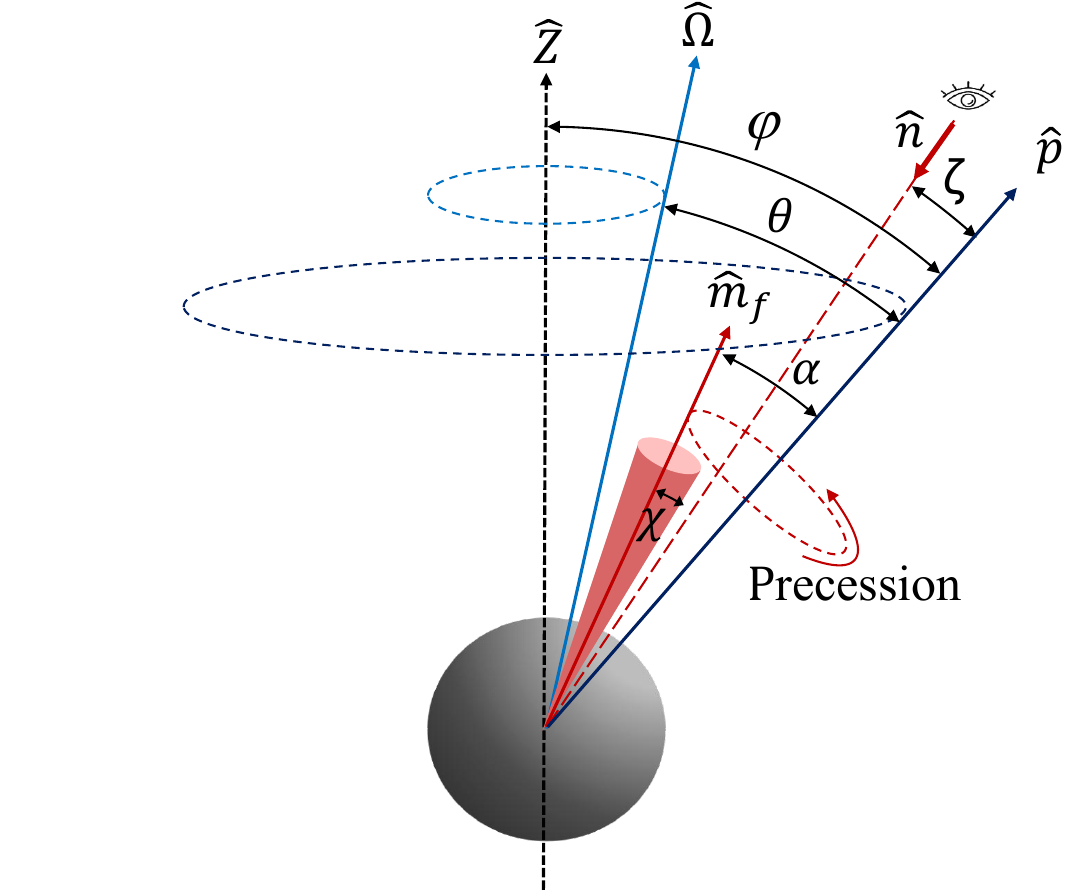}
\caption{Schematic configuration of the free precession of a magnetar in the inertial frame. $\hat p$ indicates the direction of the magnetic axis. $\hat \Omega$ corresponds to the direction of rotation angular velocity. $\hat Z$ denotes the direction of the angular momentum. $\hat m_f$ indicates the direction of the radiation beaming center, and it moves around $\hat p$ in the long term due to the precession process. $\hat n$ is the line of sight. We further define $\alpha$ as the angle between $\hat p$ and $\hat m_f$, $\chi$ as the half-opening angle of the radiation beam, $\theta$ as the obliquity, $\varphi$ as the angle between $\hat p$ and $\hat Z$, and $\zeta$ as the angle between $\hat n$ and $\hat p$.}\label{free_precession}
\end{figure}

In this section, we analyze the self-consistent evolutions of the rotation and precession of magnetars. 
We define the magnetar radius as $R$, the angular velocity as $\Omega$, the moment of inertia as $I$, the surface magnetic field as $B$, and the obliquity angle between the spin axis and magnetic axis as $\theta$. 
Since a young magnetar is deformed by the internal magnetic field, it appears free precession when its spin axis is not parallel with the deformation symmetry axis, as shown in Fig. \ref{free_precession}. 

The anisotropic pressure from the internal magnetic field could deform the magnetar into an oblate or prolate shape, which is usually described by the ellipticity parameter $\epsilon=(I_{zz}-I_{xx})/I_{xx}$, where $I_{xx}$ and $I_{zz}$ are the moments of inertia of the magnetar about $x$- and $z$-axes in body frame, respectively, and $I_{yy} = I_{xx}\neq I_{zz}$ is assumed.
For the deformation caused by the anisotropic internal magnetic pressure, the value of ellipticity parameter $\epsilon$ is given by \citep{Haskell2008,Mastrano2011}
\begin{equation}
\epsilon = 2.5\times10^{-7}\beta\left(\frac{B}{10^{14}~{\rm G}}\right)^2,
\label{epsilon}
\end{equation}
where $\beta  = (1-0.385/\Lambda)$, and $\Lambda$ is the ratio between poloidal and total magnetic energy inside the neutron star. Usually one requires $0\leqslant\Lambda\leqslant1$, where $\Lambda = 0$ and $\Lambda = 1 $ represent a star with a purely toroidal and a purely poloidal field, respectively. 
In this paper, we consider a magnetar with a mass of $M=1.4~{\rm M_\odot}$ and a radius of $R=10^6~{\rm cm}$.
If the toroidal magnetic field dominates the internal magnetic field, the neutron star would be prolate; if the polar field dominates, the neutron star would be oblate. Generally, toroidal field dominates the internal field, leading to the magnetar being prolate, so $\beta < 0$ ($\epsilon < 0$) \citep{Haskell2008,Mastrano2011,Lander2020}.
In reality, the internal magnetic field would be more complex, it usually yields $|\beta|\ll1$ in Eq.(\ref{epsilon}) \citep[see][]{Zanazzi2020}.
In the following discussion, we will discuss the different scenarios with $|\beta|$ are taken as 0.5 and 0.1 respectively. One should notice that Eq.~(\ref{epsilon}) might be invalid once the neutron star has cooled enough for its interior to be superconducting. 

Generally, the magnetar rotation is significantly affected by the magnetic dipole radiation with the torque of 
$J_{\rm E} = -T_{\rm E} B^2 \Omega^3(1+\sin^2\theta)$, where $T_{\rm E} = R^6/(4c^3)$, and the term of $(1+\sin^2\theta)$ is corrected by involving the force-free wind plasma for the charge-filled magnetosphere, leading to the torque non-zero even when the spin axis and the magnetic axis are aligned \citep{Contopoulos1999,Gruzinov2005,Timokhin2006,Spitkovsky2006}. 
In addition to the magnetic dipole radiation, due to the deformation of the magnetar, the magnetar can generate gravitational wave (GW) when the deformation axis is not parallel with the spin axis, and the torque of GW radiation is given by
$J_{\rm G} = -T_{\rm G} \epsilon^2 \Omega^5 \sin^2\theta (1+15\sin^2\theta)$ \citep{Chau1970,Melatos2000}, where $T_{\rm G} = 2GI^2/(5c^5)$. 
Generally, the torque of GW radiation $J_{\rm G}$ is too small compared with the torque of the magnetic dipole radiation $J_{\rm E}$, thus, $J_{\rm G}$ is ignored in the following calculation. The evolution equations of the angular velocity and obliquity is given by \citep{Mondal2021,Chau1970,Melatos2000,Kalita2020}
\begin{align}
I\dfrac{d \Omega}{d t}&=-T_{\rm E} B^2 \Omega^3(1+\sin^2\theta),\nonumber\\
I\Omega\dfrac{d\theta}{dt}&=-T_{\rm E}B^2\Omega^3\sin\theta\cos\theta,\label{angle_velocity}
\end{align} 
where the moment of inertia is taken as $I=10^{45} ~{\rm g~cm^2}$.
Notice that here we also ignore the contributions from the radiation and ejecta by magnetar activity (e.g., bursts, flares, etc.), because they might not significantly change the angular momentum of the magnetar during the long-term evolution due to the possible random beaming or isotropic emission/ejection.

The magnetic field evolution of a magnetar has been generally considered to be dominated by the Ohmic dissipation and Hall drift \citep{Goldreich1992}, 
and the evolution equation could be approximately written as \citep[e.g.,][]{Colpi2000,Aguilera2008,Glampedakis2011}
\begin{equation}
\frac{dB}{dt} \simeq -\frac{B}{\tau_{\rm Ohm}}-\frac{1}{B_{\rm i}}\frac{B^2}{\tau_{\rm Hall}},
\label{magnetic}
\end{equation}
where $B_{\rm i}$ is the initial magnetic field strength, and $\tau_{\rm Ohm}$ and $\tau_{\rm Hall}$ are the typical timescales of Ohmic dissipation and Hall drift decay, respectively. The typical timescales of Ohmic dissipation and Hall drift decay are taken as $\tau_{\rm Ohm} = 10^6 ~{\rm yr}$ and $\tau_{\rm Hall} = 2\times10^3 ~{\rm yr}~(B_{\rm i}/10^{15}~{\rm G})^{-1}$, respectively in the following calculations \citep[e.g.,][]{Aguilera2008,Mondal2021}.

Next, we discuss the free precession of a magnetar. Since the deformation is caused by the internal magnetic field, the deformation axis is approximately the magnetic axis. If the spin axis is not parallel to the deformation axis, the neutron star will precess around $\hat p$ axis, as shown in Fig. \ref{free_precession}, and the precession period is given by \citep[e.g.,][]{Levin2020,Zanazzi2020}
\begin{equation}
P_{\rm prec} = \frac{P}{|\epsilon \cos\theta|},\label{precession}
\end{equation}
where $P=2\pi/\Omega$ is the magnetar rotation period.
Once the initial conditions are given, the evolutions of obliquity $\theta$, surface magnetic field $B$, rotation period $P$, and precession period $P_{\rm prec}$ could be solved using the differential equations of Eq.(\ref{angle_velocity}), Eq.(\ref{magnetic}) and Eq.(\ref{precession}). 

In order to obtain the evolutions of rotation and precession, one needs know the initial obliquity. \citet{Lander2020} found that the obliquity $\theta$ tends to $\pi/2$ shortly after birth due to the viscous damping, then decreases towards zero over hundreds of years with the aligning effect of the exterior torque (magnetic dipole radiation). The early evolution of $\theta$ is mainly affected by the internal viscous damping and the magnetic dipole radiation,
\begin{equation}
\frac{d\theta}{dt} = \frac{\dot{E}_{\rm visc}}{I\epsilon\sin\theta\cos\theta\Omega^2}+\frac{\dot{E}_{\rm EM}^{(\theta)}}{I\Omega^2},\label{damping}
\end{equation}
where $\dot{E}_{\rm EM}^{(\theta)}\simeq-(R^6/4c^3)\Omega^4B^2(1+\sin^2\theta)\sin\theta\cos\theta$ is the effective electromagnetic energy-loss rate considering that $d\theta/dt$ should vanish for $\theta=0,\pi/2$ \citep{Mestel1968,Lander2020},  $\dot{E}_{\rm visc}$ is the viscous energy-loss rate dominated by shear and bulk viscosities.
Notice that $\dot{E}_{\rm visc}<0$ and $\dot{E}_{\rm EM}^{(\theta)}<0$, so that internal damping gives $d\theta/dt<0$ for the oblate deformation ($\epsilon>0$) and $d\theta/dt>0$ for the prolate deformation ($\epsilon<0$). The typical timescale for the electromagnetic term is
\begin{align}
\tau_{\rm EM}\sim\frac{4Ic^3}{R^6\Omega^2B^2}\simeq2.7\times10^3~{\rm s}\left(\frac{B}{10^{15}~{\rm G}}\right)^{-2}\left(\frac{P}{1~{\rm ms}}\right)^2.
\end{align}
The viscous energy-loss rate is mainly dominated by bulk viscosities, and the typical timescale is estimated by \citep{Lander2018}
\begin{align}
&\tau_{\rm visc}=8~{\rm s}\left(\frac{T}{10^{10}~{\rm K}}\right)^6\left(\frac{P}{1~{\rm ms}}\right)^4\left(\frac{B}{10^{15}~{\rm G}}\right)^{-2}\nonumber\\
&\times\left[1+0.02\left(\frac{T}{10^{10}~{\rm K}}\right)^{-12}\left(\frac{P}{1~{\rm ms}}\right)^{-2}\left(\frac{B}{10^{15}~{\rm G}}\right)^{4}\right].
\end{align} 
We set the corresponding time-scales equal, $\tau_{\rm EM}\sim\tau_{\rm visc}$, and obtained the critical rotation period is about $P_{\rm cr}\sim18~{\rm ms}$ for the about typical parameters. Thus, the critical timescale is $\tau_{\rm cr}\sim10$ days. For a neutron star with age $\lesssim\tau_{\rm cr}$, the evolution of the obliquity $\theta$ is dominated by bulk viscosities. In particular, for the prolate deformations with $\epsilon<0$, one has $d\theta/dt>0$, leading to the obliquity $\theta$ tending to $\pi/2$ shortly after birth.
Since the critical timescale $\tau_{\rm cr}$ is small compared with the whole age of a neutron star, in the following calculation, the initial value of the obliquity could be set close to $\pi/2$. In the following numerical calculations, we take it as $\pi/2-\delta$ with $\delta\sim0.1$. 
We take the initial period $P_{\rm i}$ from $5~{\rm ms}$ to $80 ~{\rm ms}$ and take the initial magnetic field $B_{\rm i}$ from $10^{14}{\rm G}$ to $10^{16} {\rm G}$. For the parameter $\beta$ in the ellipticity equation Eq.(\ref{epsilon}), we take $|\beta|=0.5$ and $|\beta|=0.1$.

\begin{figure*}
\centering
\includegraphics[width=7.5cm]{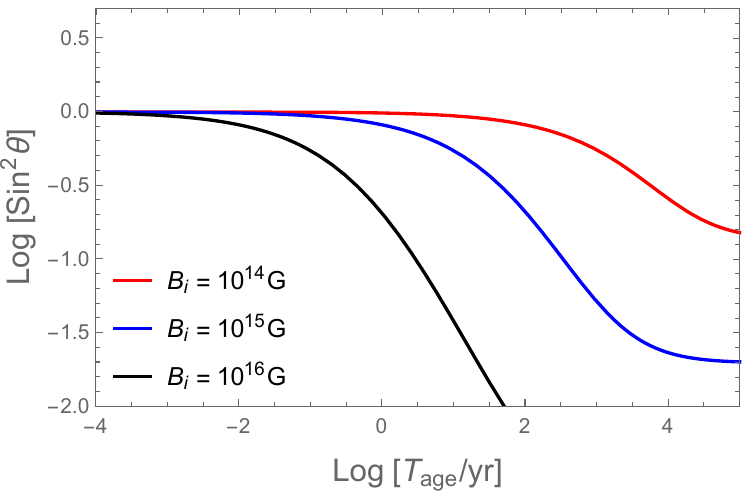}
\includegraphics[width=7.5cm]{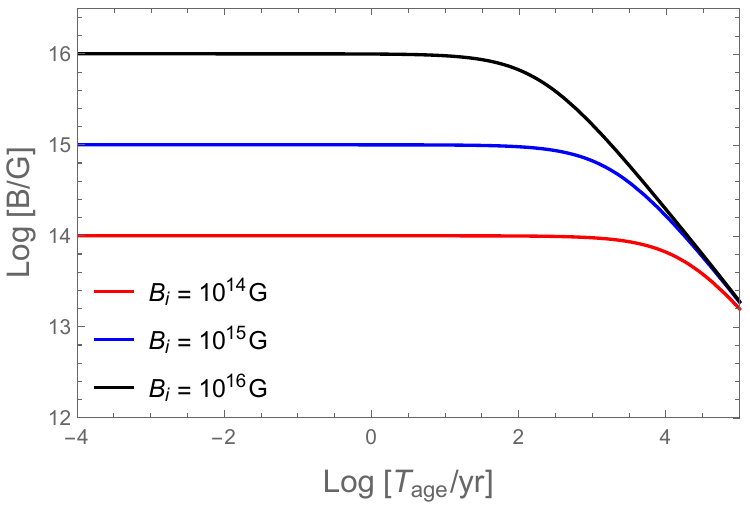}
\includegraphics[width=7.5cm]{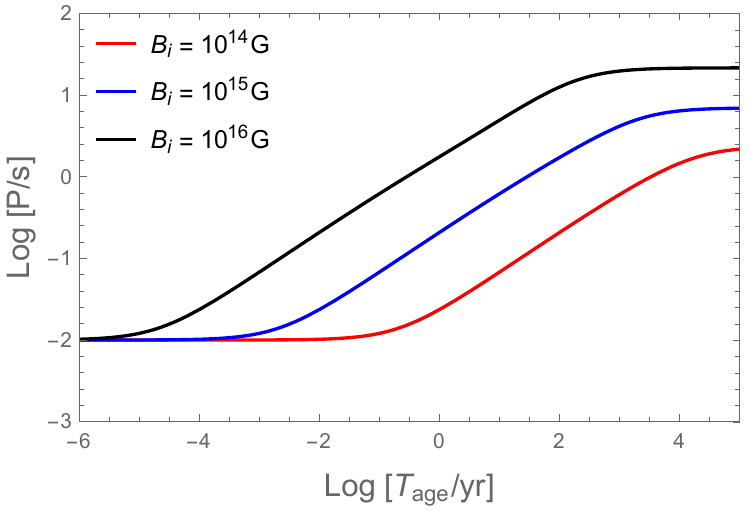}
\includegraphics[width=7.5cm]{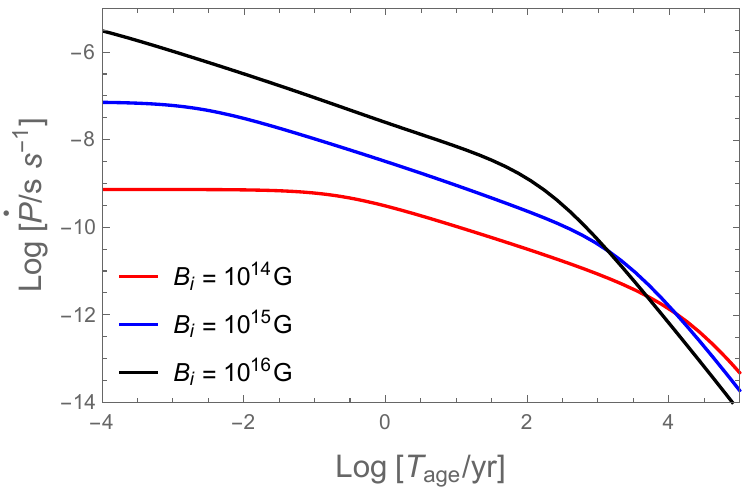}
\caption{The evolutions of obliquity $\theta$, surface magnetic field $B$, rotation period $P$ and period derivative $\dot P$. The initial parameters are set to $B_{\rm i} = 10^{14}, 10^{15},  10^{16} ~{\rm G}$, $P_{\rm i} = 10 ~{\rm ms}$ and $|\beta| = 0.1$.}\label{quantities_diverse_B}
\end{figure*}

\begin{figure*}
\centering
\includegraphics[width=7.5cm]{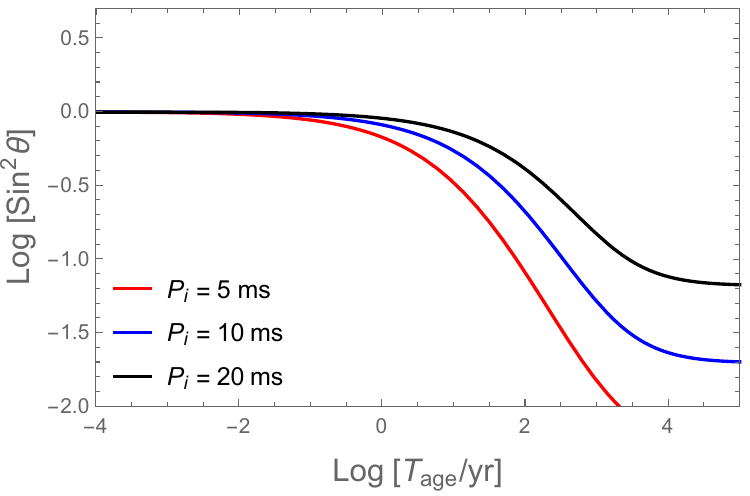}
\includegraphics[width=7.5cm]{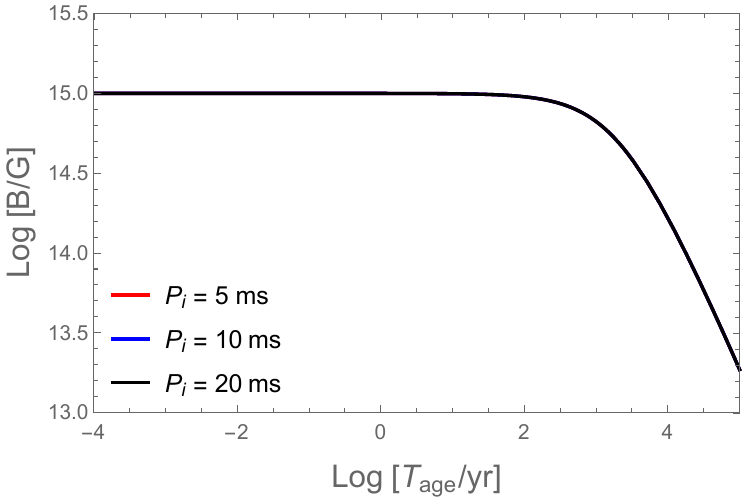}
\includegraphics[width=7.5cm]{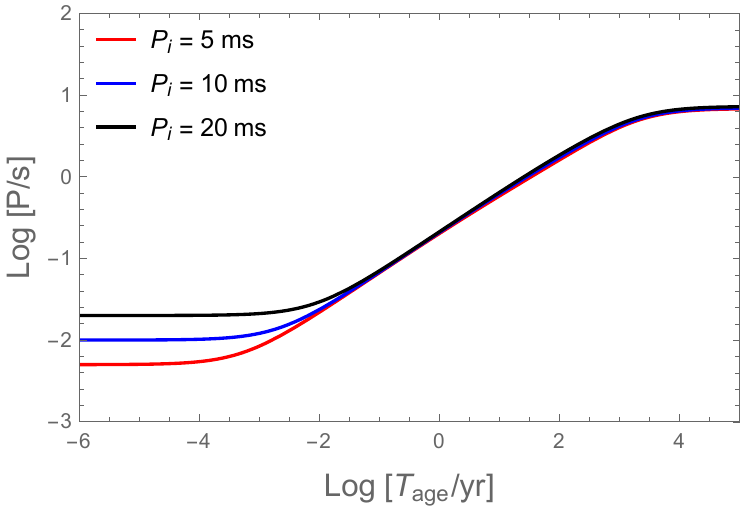}
\includegraphics[width=7.5cm]{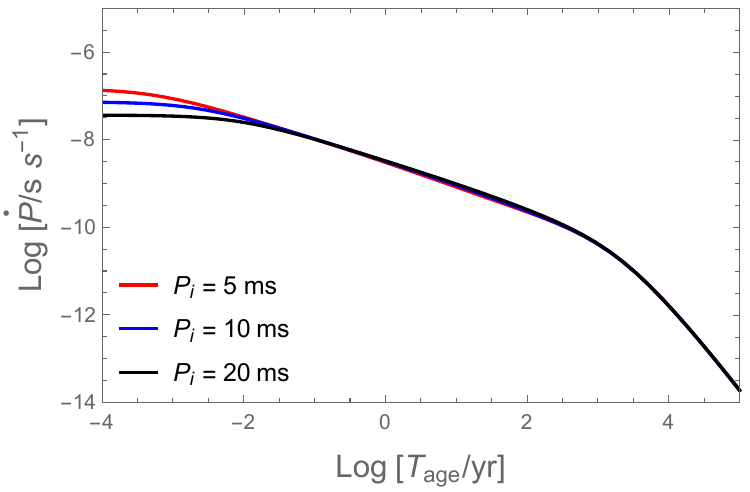}
\caption{Same as Fig. \ref{quantities_diverse_B}. The initial parameters are set to $P_{\rm i} = 5, 10, 20 ~{\rm ms}$, $B_{\rm i} = 10^{15} ~{\rm G}$ and $|\beta| = 0.1$.}\label{quantities_diverse_P}
\end{figure*}

Fig. \ref{quantities_diverse_B} shows the evolutions of the obliquity $\theta$, the surface magnetic field $B$, the spin period $P$ and the spin periodic derivative $\dot{P}$ with different initial surface magnetic fields of $B_{\rm i} = 10^{14}, 10^{15}, 10^{16} {\rm G}$. The initial period is taken as $P_{\rm i} = 10 ~{\rm ms}$, and the deformation parameter is taken as $|\beta| = 0.1$. 
The obliquity decreases over a long term, and the stronger the initial surface magnetic field, the faster the obliquity evolution. 
The spin period increases during the long term and trends to a constant value at the later stage. One should notice that the evolution of the magnetic field significantly affects the spin evolution for magnetars. As shown in Eq.(\ref{magnetic}), the magnetic field of a magnetar would significantly decay after $\sim 10^3~{\rm yr}~(B_{\rm i}/10^{15}~{\rm G})^{-1}$ due to the Hall drift. This is the reason why the spin period tends to a constant value at late times.
For the above typical parameters, the surface magnetic field decreases significantly after $\gtrsim(100-10^4)~{\rm yr}$. 
Similarly, in Fig. \ref{quantities_diverse_P}, we show the evolutions of $\theta$, $B$, $P$ and $\dot P$ with different initial spin periods of $P_{\rm i} = 5,10,20 ~{\rm ms}$. The initial surface magnetic field is $B_{\rm i} = 10^{15} {\rm G}$, and the deformation parameter is $|\beta| = 0.1$. One can see that the faster the initial spin, the faster the obliquity evolution, meanwhile, the final obliquity depends on the initial spin period $P_{\rm i}$. The evolution of the surface magnetic field and the late evolution of the spin period are almost independent of $P_{\rm i}$.


\begin{figure*}
\centering
\includegraphics[width=7.5cm]{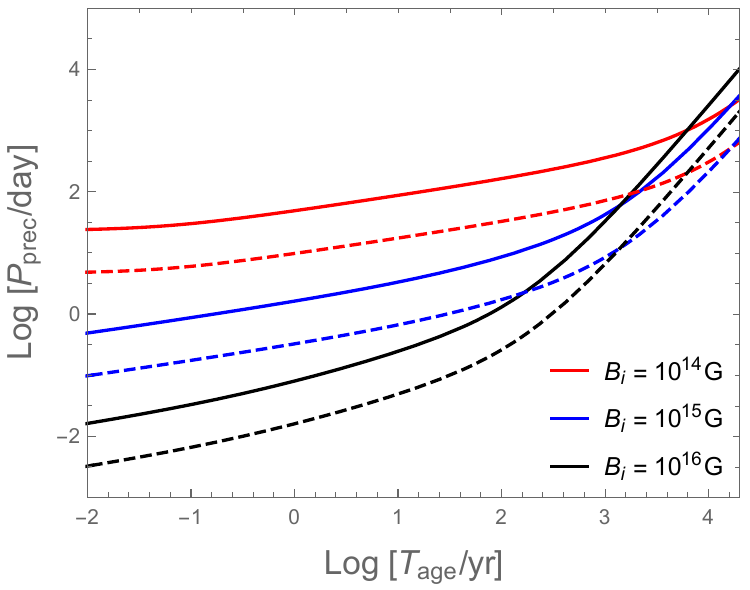}
\includegraphics[width=7.5cm]{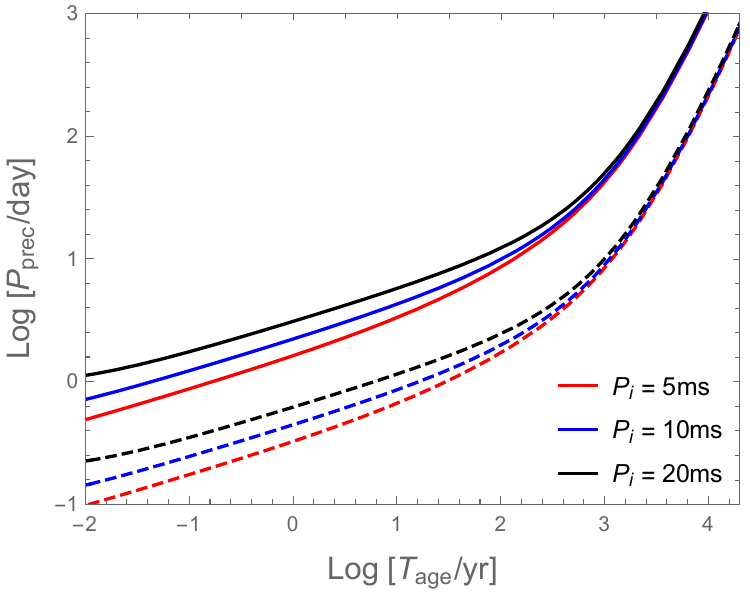}
\caption{The evolution of the precession period $P_{\rm prec}$ for different initial magnetic fields (left panel) and different initial periods (right panel) with $|\beta|=0.1$ (solid line) and $|\beta|=0.5$ (dashed line). In the left panel, $P_{\rm i}=5~{\rm ms}$, and in the right panel, $B_{\rm i}=10^{15}~{\rm G}$.}\label{precession_period}
\end{figure*}

Once the evolutions of $P$, $\theta$, and $B$ are obtained, one can calculate the evolution of the free precession period $P_{\rm prec}$ based on Eq.(\ref{precession}). In Fig. \ref{precession_period}, the left panel and right panel correspond to $P_{\rm prec}$ evolution for different initial surface magnetic fields and different initial periods, respectively. In the left panel, $P_{\rm i}=5~{\rm ms}$ is taken, and in the right panel, $B_{\rm i}=10^{15}~{\rm G}$ is taken. The solid and dashed lines corresponds to $|\beta|=0.1$ and $|\beta|=0.5$, respectively.
We find that the evolution of the precession period satisfies a broken power law in the long term. The evolution of the precession period increases slowly in the early stage and increases fast in the later stage. We define the broken timescale as the critical value between these two stages, and the broken timescale mainly depends on $B_{\rm i}$ and is almost independent of $P_{\rm i}$ and $|\beta|$ as shown in Fig. \ref{precession_period}.
For the above typical parameters, the evolution of $P_{\rm prec}$ increase faster after $\sim 1000~{\rm yr}$ for $B_{\rm i}=10^{15}~{\rm G}$. 
The larger the initial surface magnetic field or the faster the initial spin, the smaller the precession period. Besides, a larger deformation parameter $|\beta|$ would also cause a smaller precession period. 

\section{Constraint on the ages of magnetars with periodic activities}\label{sec3}

\begin{figure*}
\centering
\includegraphics[width=7.5cm]{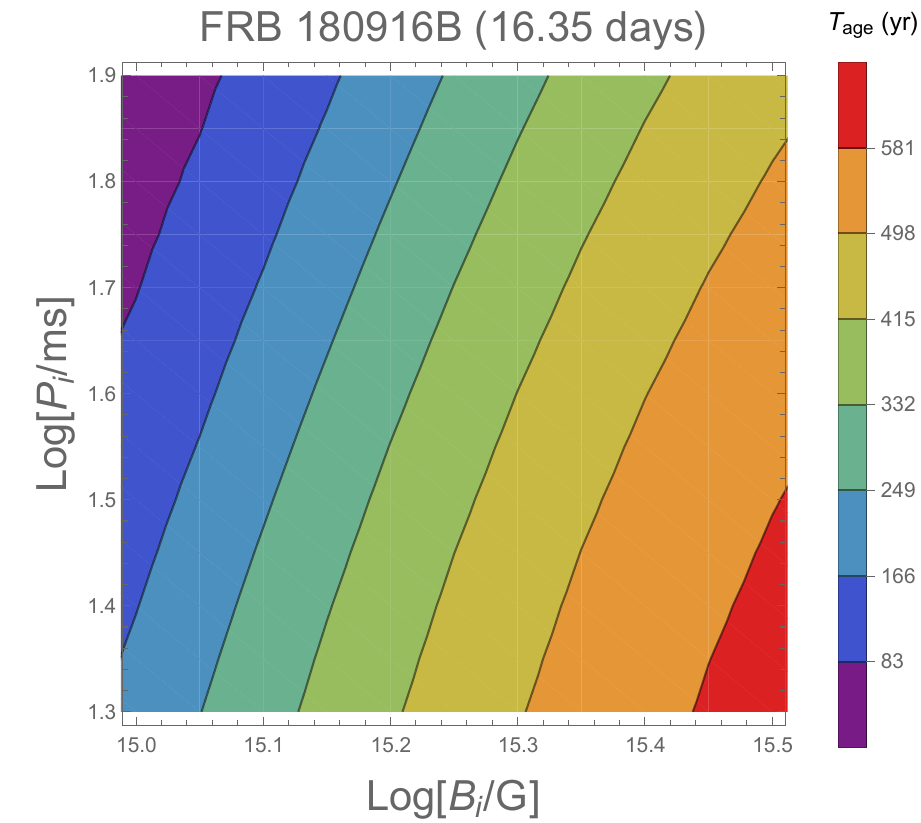}
\includegraphics[width=7.5cm]{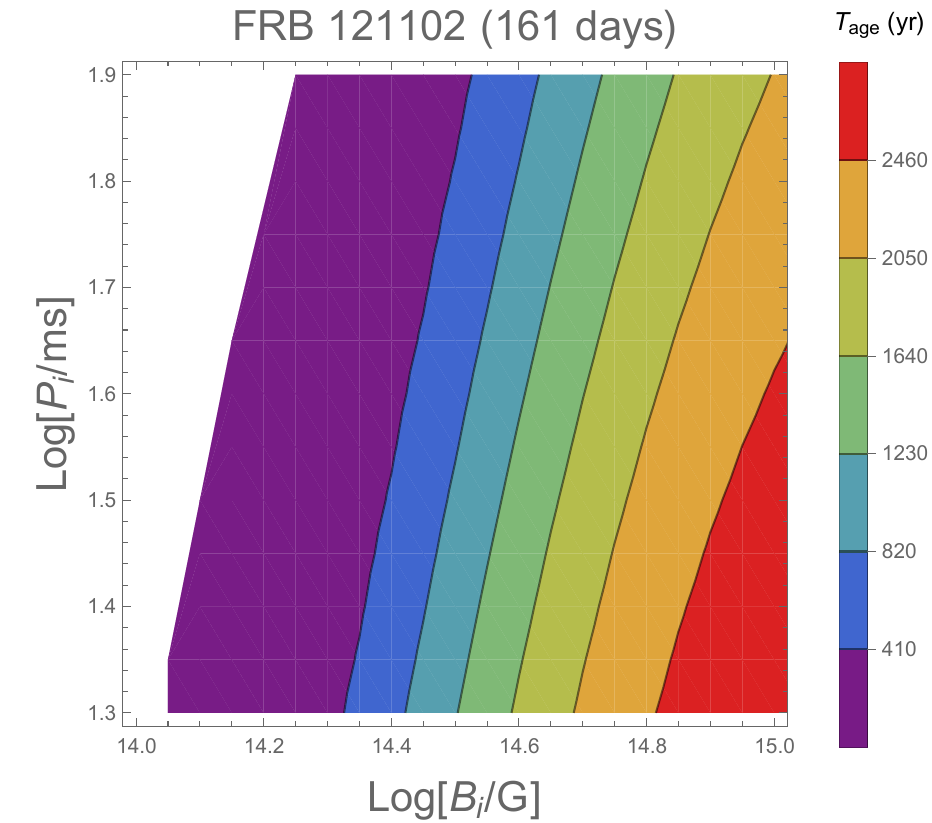}
\caption{Contour diagram of the magnetar ages for different sources with the precession period of 16.35 days for FRB 180916B, 161 days for FRB 121102. We take the initial period $P_{\rm i}$ with range $(20~{\rm ms},80~{\rm ms})$ and the initial magnetic field $B_{\rm i}$ with range $(10^{14} ~{\rm G},10^{16} ~{\rm G})$, and $|\beta|=0.1$. In this figure, the white regions represent no solution for the given precession periods.}\label{contour01}
\end{figure*}

\begin{figure*}
\centering
\includegraphics[width=7.5cm]{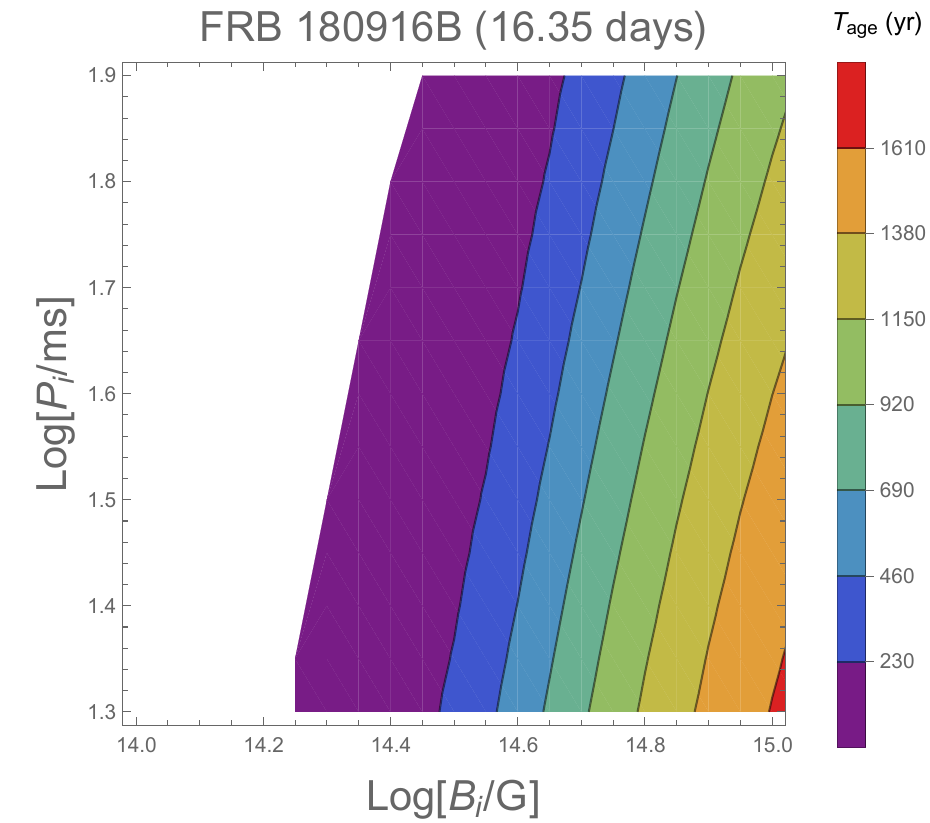}
\includegraphics[width=7.5cm]{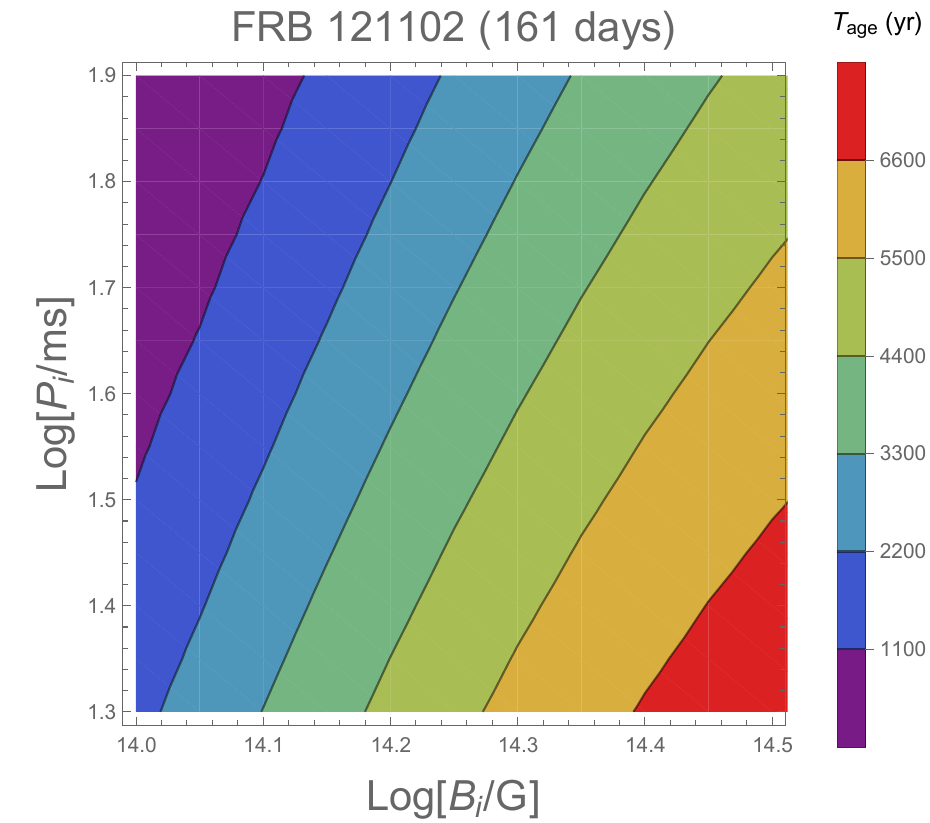}
\caption{Same as Fig. \ref{contour01} but with $|\beta|=0.5$.}\label{contour05}
\end{figure*}

Based on the evolution of the precession period, we can constrain the ages of magnetars that show periodic activities.
Recently, some FRB repeaters and SGRs were found to show (possible) periodic activities, such as FRB 180916B, FRB 121102, SGR 1806-20, and SGR 1935+2154 \citep{ChimeFrbCollaboration2020,Rajwade2020,ZhangGQ2021,Zou2021,Xie2022}. 
Assuming that their periodicities are due to the free precession of a magnetar \citep{Levin2020,Zanazzi2020,LiDongZi2020}, one can constrain the ages of these sources, as shown in Fig. \ref{contour01} ($|\beta|=0.1$) and Fig. \ref{contour05} ($|\beta|=0.5$) in the parameter space of $(B_{\rm i},P_{\rm i})$.

FRB 180916B was found to be localized in a spiral galaxy at redshift $z = 0.0338$ \citep{Tendulkar2021}. This source appears a 16.35-day periodic activity, and its activity window is about 5 day \citep{ChimeFrbCollaboration2020}, and the activity window was found to be frequency-dependent \citep{Pastor-Marazuela2021,Bethapudi2022}. Taking $P_{\rm prec}=16.35~{\rm day}$, the magnetar age could be constrained in the parameter space $(B_{\rm i},P_{\rm i})$, as shown in the subpanels of Fig. \ref{contour01} and Fig. \ref{contour05}, respectively. The observed $P_{\rm prec}=16.35~{\rm day}$ requites that the magnetar age ranges from $\sim 100~{\rm yr}$ to $\sim 2000~{\rm yr}$ for the initial period with $P_{\rm i}\sim(20~{\rm ms}, 80~{\rm ms})$, initial surface magnetic field with $B_{\rm i}\sim(10^{14}~{\rm G}, 10^{16}~{\rm G})$, and deformation parameter $|\beta|\sim(0.1, 0.5)$.  Compared with the subpanels of Fig. \ref{contour01} and Fig. \ref{contour05}, one can find that the more significant the deformation (the larger $|\beta|$ or $|\epsilon|$), the longer the constrained magnetar age. The reason is that for a given observed precession period, the more significant the deformation (the larger $|\beta|$ or $|\epsilon|$), the longer the magnetar age, as shown in Fig. \ref{precession_period}.

FRB 121102, the first known repeater, was localized in a star-forming dwarf galaxy at $z = 0.19273$ \citep{Chatterjee2017,Tendulkar2017}. This source has a possible $\sim 161$-day periodic activity with 60\% duty cycle \citep{Rajwade2020,Cruces2021}.
Taking $P_{\rm prec}=161~{\rm day}$, the magnetar age could be constrained in the parameter space $(B_{\rm i}, P_{\rm i})$, as shown in the subpanels of Fig. \ref{contour01} and Fig. \ref{contour05}, respectively. The observed $P_{\rm prec}=161~{\rm day}$ requites that the magnetar age ranges from $\sim 400~{\rm yr}$ to $\sim 7000~{\rm yr}$ for the initial period with $P_{\rm i}\sim(20~{\rm ms}, 80~{\rm ms})$, initial surface magnetic field with $B_{\rm i}\sim(10^{14}~{\rm G}, 10^{16}~{\rm G})$, and deformation parameter $|\beta|\sim(0.1, 0.5)$.

In addition to FRB repeaters, some SGRs also show possible periodic X-ray activities. A Galactic magnetar, SGR 1935+2154, has been active for a few years. It is very interesting that FRB 200428 was detected from SGR 1935+2154 associated with one X-ray burst, which implies that at least a part of FRBs originate from magnetars. Recently, SGR 1935+2154 was found to exhibit a possible period of $127$ day in X-ray activity \citep{Xie2022}. We consider that the periodic activities are due to the free precession of magnetar. Combining the observation of $P = 3.245 ~{\rm s}$ and $\dot{P} = 1.43\times 10^{-11}{\rm s ~s^{-1}}$ \citep{Olausen2014}, we can obtain the age, initial magnetic field, and $|\beta|$ of this magnetar for a given initial period. For $P_{\rm i}=5~{\rm ms}$, one has $T_{\rm age}=2136.6~{\rm yr}$, $B_{\rm i}=5.66\times10^{14}~{\rm G}$, and $|\beta|=0.0964$. For $P_{\rm i}=10~{\rm ms}$, one has $T_{\rm age}=2103~{\rm yr}$, $B_{\rm i}=5.37\times10^{14}~{\rm G}$, and $|\beta|=0.105$. 

Another active Galactic magnetar, SGR 1806-20, also shows a possible $398.2$-day periodic activity \citep{ZhangGQ2021}. Combining the observation of $P = 7.55 ~{\rm s}$ and $\dot{P} = 49.5\times 10^{-11}{\rm s ~s^{-1}}$ \citep{Olausen2014}, we can obtain the age, initial magnetic field, and $|\beta|$ of this magnetar.
For $P_{\rm i}=5~{\rm ms}$, one has $T_{\rm age}=172.6~{\rm yr}$, $B_{\rm i}=4.40\times10^{15}$, and $|\beta|=0.000865$. For $P_{\rm i}=10~{\rm ms}$, one has $T_{\rm age}=169.9~{\rm yr}$, $B_{\rm i}=4.33\times10^{15}~{\rm G}$, and $|\beta|=0.000859$.

\section{Observable probability and active window of FRBs}\label{sec4}

\begin{figure}
\centering
\includegraphics[width=7.5cm]{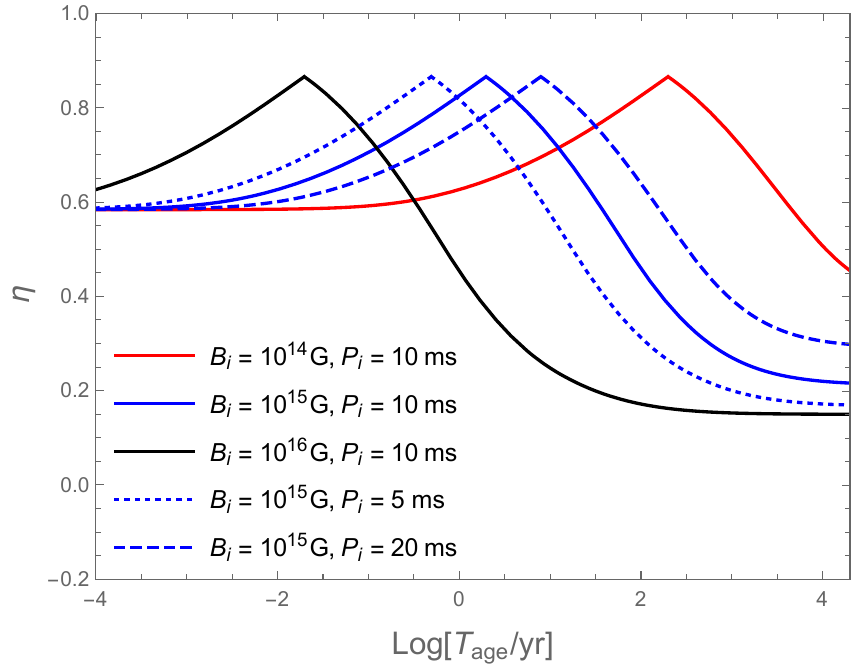}
\caption{The evolution carve of observable probability $\eta$. We take $|\beta| = 0.1$, and $\alpha = \pi/9$, $\chi = \pi/18$. Here we consider that the bursts are observable as long as they are within the radiation beam and only the geometrical effects are considered here.}\label{observable_probability} 
\end{figure}

Due to the magnetar rotation, precession, and beaming radiation, one can only observe the radiation at some certain directions of the line of sight, as shown in Fig. \ref{free_precession}. In particular, for FRB repeaters, we assume that radio bursts are generated in the magnetosphere as proposed in some magnetosphere models \citep[e.g.,][]{YangYuanPei2018,Yang2021,Kumar2020,LuWenBin2020}. 
When the magnetosphere is triggered by the crust fracturing of a magnetar, Alfv\'en waves are produced and propagate along the field lines, and a charge starvation region is formed in the magnetosphere due to the parallel electric fields induced by the Alfv\'en waves. Coherent radio bursts are produced by coherent curvature radiation by charged bunches \citep{YangYuanPei2018,Kumar2020,LuWenBin2020} or coherent plasma radiation due to nonuniform pair production across magnetic field lines \citep{Philippov2020,Yang2021,Mahlmann2022}.  
We notice that for the free precession of a biaxial neutron star if the radiation region is perfectly axisymmetric with respect to the deformation axis of the moment of inertia tensor, the slow precession of the neutron star about this axis would not produce any observable changes. Therefore, there has to be some non-axisymmetry with respect to the deformation axis to produce observable changes. In the following discussion, we assume that the beaming direction of the FRB emission region deviates from the deformation axis, as shown in Fig \ref{free_precession}.
We assume that the FRBs are emitted from a beaming region with an angle $\alpha$ from the magnetic axis $\hat p$. The half-opening angle of the radiation beam is $\chi$. During the precession process, the beaming center direction $\hat m_f$ moves around $\hat p$ in the long term. 
Due to the rotation and the precession, the radiation beam will sweep across a solid angle that corresponds to a region where the radiation is observable. For $\pi/2-(\alpha+\chi)\le\varphi$, the solid angle is $\Psi = 2[2\pi(1-\cos(\pi/2))-2\pi(1-\cos(\varphi-(\alpha+\chi)))]$, where the factor of $2$ is involved due to the beaming radiation from the two opposite poles considered, $\varphi$ is the angle between $\hat p$ and $\hat Z$. For $(\alpha+\chi)\le\varphi\le\pi/2-(\alpha+\chi)$, the solid angle is $2[2\pi(1-\cos(\varphi+\alpha+\chi))-2\pi(1-\cos(\varphi-(\alpha+\chi)))]$. For $(\alpha-\chi)\le\varphi\le(\alpha+\chi)$, the solid angle is $2[2\pi(1-\cos(\varphi+\alpha+\chi))]$. For $\varphi\le(\alpha-\chi)$, the solid angle is $2[2\pi(1-\cos(\varphi+\alpha+\chi))-2\pi(1-\cos(\alpha-\chi-\varphi))]$. In summary, the solid angle that corresponds to the radiation is observable is given by
\begin{equation}
\Psi = \begin{cases}4\pi\cos[\varphi-(\alpha+\chi)],&\frac{\pi}{2}-(\alpha+\chi)\le\varphi\\
8\pi\sin\varphi\sin(\alpha+\chi),&(\alpha+\chi)\le\varphi\le\frac{\pi}{2}-(\alpha+\chi)\\
4\pi[1-\cos(\varphi+\alpha+\chi)],&(\alpha-\chi)\le\varphi\le(\alpha+\chi)\\
8\pi\sin\alpha\sin(\varphi+\chi),&\varphi\le(\alpha-\chi)\end{cases}.\label{solid_angle}
\end{equation}
For a certain source, the observation direction is uniformly randomly distributed in space. Thus, the corresponding observable probability is given by
\begin{equation}
\eta = \frac{\Psi}{4\pi} =  \begin{cases}\cos[\varphi-(\alpha+\chi)],&\frac{\pi}{2}-(\alpha+\chi)\le\varphi\\
2\sin\varphi\sin(\alpha+\chi),&(\alpha+\chi)\le\varphi\le\frac{\pi}{2}-(\alpha+\chi)\\
1-\cos(\varphi+\alpha+\chi),&(\alpha-\chi)\le\varphi\le(\alpha+\chi)\\
2\sin\alpha\sin(\varphi+\chi),&\varphi\le(\alpha-\chi)\end{cases}.\label{probability}
\end{equation}
Notice that the observable probability only involves the geometrical effect, and other effects (e.g., burst frequency, flux threshold, etc.) are not considered here. We define the angular momentum of the magnetar as $L$ for a free body, and the direction of $L$ is $\hat Z$ shown in Fig. \ref{free_precession}. In general, the direction of the rotation angular velocity $\Omega$ is not collinear to the direction of $L$. The components of $\Omega$ are given by
\begin{equation}
\Omega_{xx}~=~\frac{L}{I_{xx}}\sin\varphi,\label{}
\end{equation}
\begin{equation}
\Omega_{zz}~=~\frac{L}{I_{zz}}\cos\varphi,\label{}
\end{equation}
where $\Omega_{xx}$ and $\Omega_{zz}$ are the projections of $\Omega$ onto x-axis and z-axis in the body frame. On the other hand, the obliquity angle is given by
\begin{equation}
\tan\theta~=~\frac{\Omega_{xx}}{\Omega_{zz}}.\label{}
\end{equation}
leading to,
\begin{equation}
\tan\theta~=~\frac{I_{zz}}{I_{xx}}\tan\varphi~=~(1+\epsilon)\tan\varphi.\label{}
\end{equation}
The angle $\varphi$ evolves due to the evolution of the obliquity $\theta$ and the ellipticity parameter $\epsilon$, leading to the evolution of the observable probability $\eta$.
Fig. \ref{observable_probability} shows the relation between the observable probability $\eta$ and the magnetar age $T_{\rm age}$. 
The red, blue, and black solid lines denote the relations with $P_{\rm i} = 10~{\rm ms}$ and $B_{\rm i} = 10^{14}, 10^{15}, 10^{16} {\rm G}$, respectively. The dotted, solid, and dashed blue lines denote the relations with $B_{\rm i} = 10^{15}{\rm G}$ and $P_{\rm i} = 5, 10, 20~{\rm ms}$. We take $|\beta| = 0.1$, and $\alpha = \pi/9$, $\chi = \pi/18$. 
We can see that for a given magnetar with an initial magnetic field of $B_{\rm i}$ and a rotation period of $P_{\rm i}$, the observable probability $\eta$ increases with the magnetar age $T_{\rm age}$ in the early stage. The reason is that the solid angles swept by the two radiation beams near the magnetic poles are overlapped at $\theta\sim\varphi\sim\pi/2$ and the total solid angle increases as the $\varphi$ becomes smaller. In the later stage, the two solid angles swept by the two radiation beams are separated and the observable probability $\eta$ decreases with the magnetar age $T_{\rm age}$, because the solid angle swept by one radiation beam becomes smaller as the $\varphi$ becoming smaller. 
We notice that only the geometrical effect is considered in the above discussion. Furthermore, since young magnetars with stronger magnetic fields and faster spins tend to produce more frequent repeated bursts triggered by the crust quake of neutron stars \citep{Yang2021,Li2022}, the probability of FRBs observed in the early stage might be much larger than that in the later stage.

\begin{figure}
\centering
\includegraphics[width=7.5cm]{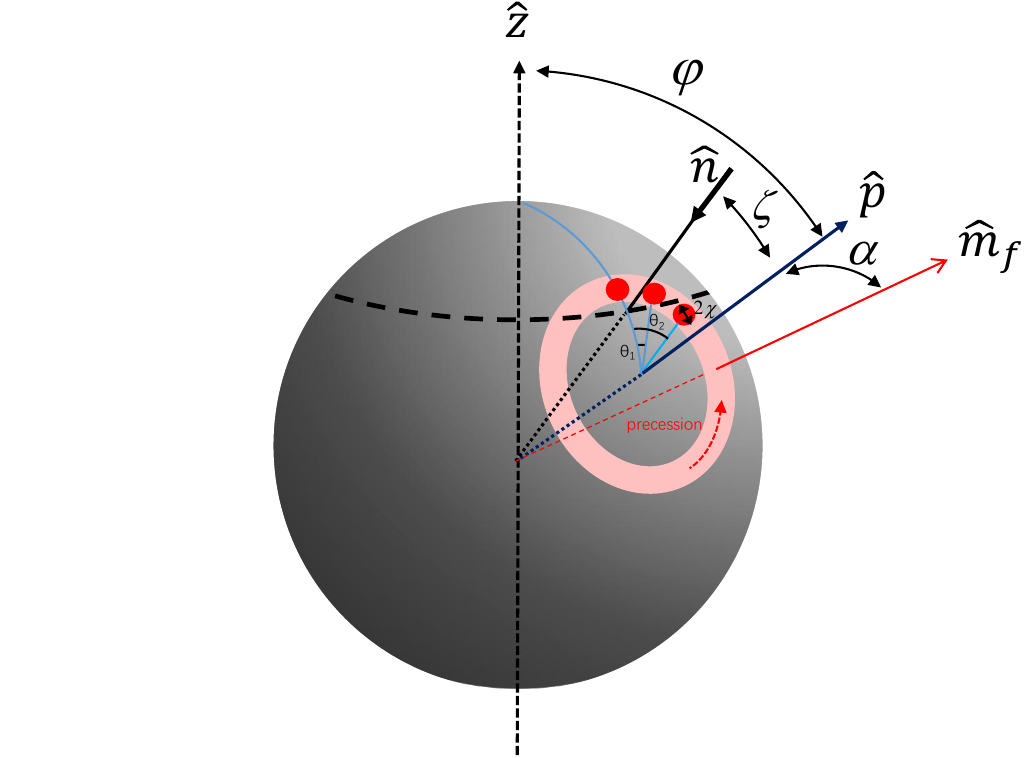} 
\caption{Schematic configuration of the motion process of a radiation beam in the rotation frame where $\hat Z$ and $\hat p$ are fixed. 
The dashed line denotes the line of sight. The small circle denotes the radiation beam, and the ring band denotes the region swept by the radiation beam due to the precession in this frame.
$\varphi$ is the angle between $\hat p$ and $\hat Z$. $\zeta$ is the angle between the line of sight and $\hat p$. $\alpha$ is the angle between $\hat p$ and $\hat m_f$. $\chi$ is the half-opening angle of the radiation beam. $\theta_1 = \arccos[(\cos(\varphi-\zeta-\chi)-\cos\varphi\cos\alpha)/(\sin\varphi\sin\alpha)]$, and $\theta_2 = \arccos[(\cos(\varphi-\zeta+\chi)-\cos\varphi\cos\alpha)/(\sin\varphi\sin\alpha)]$.}\label{free_precession2}
\end{figure}

Based on the above model, we further analyze the active window distribution of the periodic activities of an FRB repeater as follows. We consider the motion process of a radiation beam in the rotation frame where the line of sight rotates around $\hat Z$ axis and the radiation beam processes around $\hat p$ axis, as shown in Fig. \ref{free_precession2}. 
We emphasize that the rotation frame defined here is not the same as the body frame described in literature \citep[e.g.,][]{LiDongZi2020}. In the rotation frame, $\hat Z$ and $\hat p$ are considered to be fixed.
The trajectory of the line of sight in the rotation frame is shown in the dashed line. The red circle denotes the radiation beam, and the pink ring band denotes the region swept by the radiation beam due to the precession in the rotation frame.
Thus, the active window corresponds to the period when the radiation beam crosses over the line of sight.

\begin{figure*}
\centering
\includegraphics[width=7.5cm]{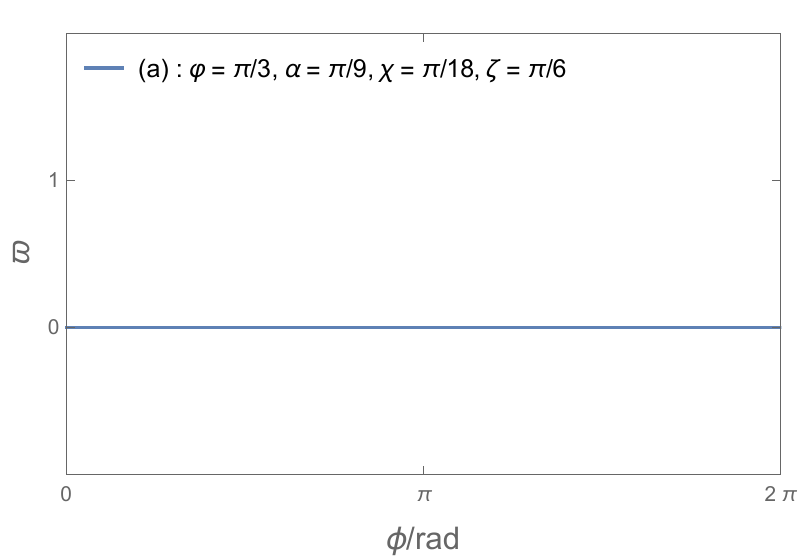}
\includegraphics[width=7.5cm]{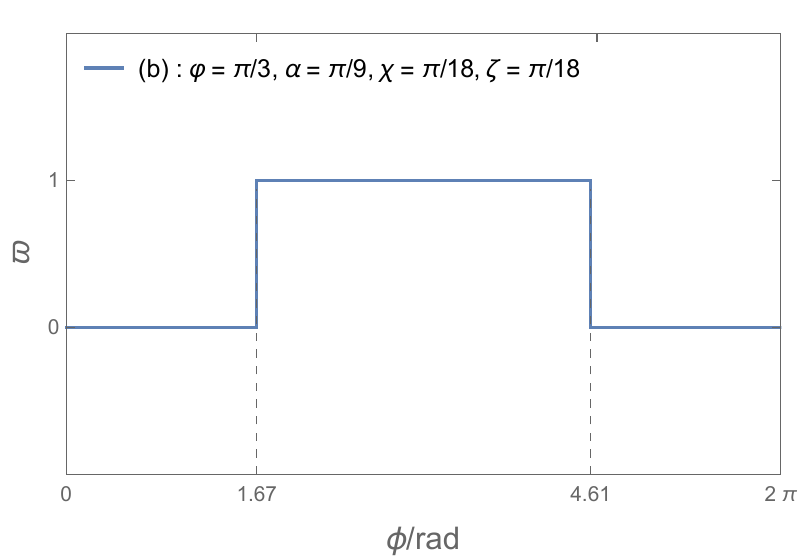}
\includegraphics[width=7.5cm]{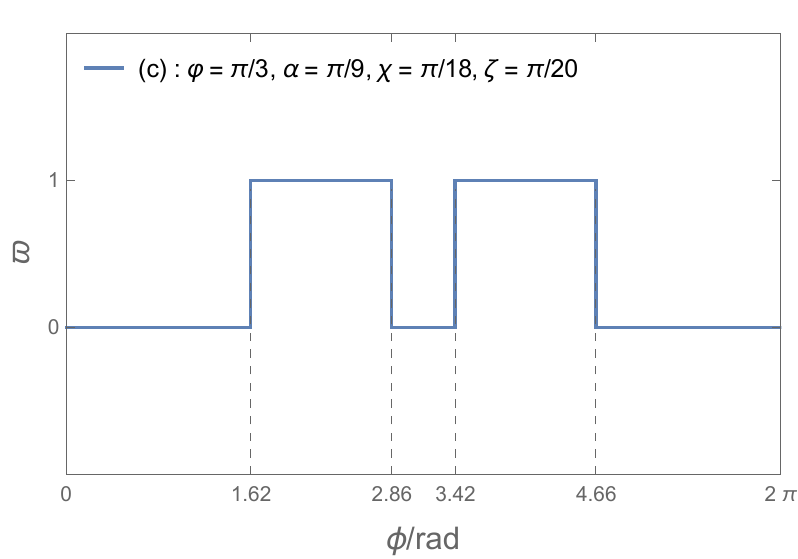}
\includegraphics[width=7.5cm]{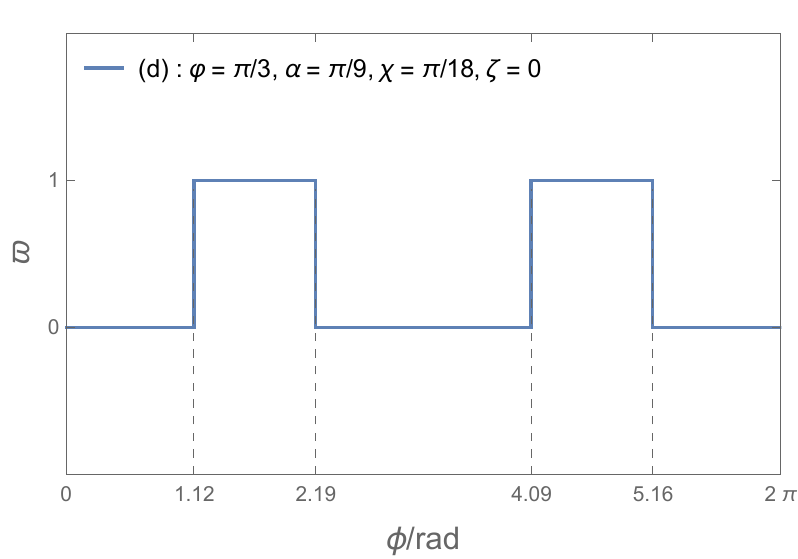}
\caption{The phase distribution of the activity window in one precession period. 
$\varpi=1$ corresponds to the active window and $\varpi=0$ corresponds to the nonactive window.
Different panels correspond to different directions of the line of sight with $\zeta=\pi/6,\pi/18,\pi/20,0$, respectively. $\varphi = \pi/3$, $\alpha = \pi/9$ and $\chi = \pi/18$ are taken here for example.
}\label{case}
\end{figure*}

In Fig. \ref{free_precession2}, the radiation beam moves around $\hat p$ due to the precession.
$\theta_2$ and $\theta_1$ correspond to the angles for entering and exiting the active window, respectively, which are given by
\begin{align}
\theta_1 &=\arccos\frac{\cos(\varphi-\zeta-\chi)-\cos\varphi\cos\alpha}{\sin\varphi\sin\alpha},\nonumber\\
\theta_2 &=\arccos\frac{\cos(\varphi-\zeta+\chi)-\cos\varphi\cos\alpha}{\sin\varphi\sin\alpha}.\label{theta12}
\end{align}
There are three cases for the distribution of the active window: 1) If the line of sight (dashed line) passes through the pink ring band only once, i.e., $(\alpha-\chi)\le\zeta\le(\alpha+\chi)$, there is only one active window in one precession period and the duty cycle is $(2\theta_2)/2\pi$. 2) If the line of sight passes through the pink ring band two times, i.e., $\zeta\le(\alpha-\chi)$, there are two active windows in one precession period and the duty cycle is $(2/2\pi)\left(\theta_2-\theta_1\right)$. 3) If the line of sight does not pass through the pink ring region, there is no active window. In summary, the duty cycle is given by

\begin{equation}
\Pi =\begin{cases}\frac{1}{\pi}\theta_2,&(\alpha-\chi)\le\zeta\le(\alpha+\chi)\\
\frac{1}{\pi}\left(\theta_2-\theta_1\right),&\zeta\le(\alpha-\chi)\\
0,&(\alpha+\chi)\le\zeta\end{cases}.\label{duty}
\end{equation}
In particular, for $\alpha\le\chi$, there is always one active window in one precession period once $\zeta\le\alpha+\chi$. We also notice that the above equation is valid for $P \ll P_{\rm prec}$, which could be naturally satisfied in this work.
In Fig. \ref{case}, we plot the distribution of the active window in one precession period for different directions of the line of sight directions, and $\varphi = \pi/3$, $\alpha = \pi/9$ and $\chi = \pi/18$ are taken for example.
We define the parameter $\varpi$ to describe whether the FRB source is in the active window: $\varpi=1$ corresponds to the active window and $\varpi=0$ corresponds to the nonactive window.
The initial phase corresponds to the one when the radiation beam direction $\hat m_f$ is farthest from the line of sight. For given $\varphi$, $\alpha$, and $\chi$, the distribution of the active window mainly depends on the direction of the line of sight $\zeta$, as pointed out in the above discussion. This result suggests that there are one or two active windows in one precession period.

\section{Conclusion and discussion}\label{sec5}

Some FRB repeaters and SGRs appear (possible) periodic activities with periods of tens to hundreds days, which might originate from the free precession of a young magnetar. The young magnetar is deformed by the anisotropic internal magnetic pressure, and the free precession is generated when the magnetic axis is not parallel with the spin axis. In this work, we analyze the self-consistent rotation evolution of deformed magnetars and calculate the corresponding precession evolution. We find that the obliquity evolves from $\pi/2$ to a small value over hundreds of years, and the spin period tends to be a constant finally. Due to the evolution of the spin, the obliquity, and the surface magnetic field, the precession period increases faster after a broken timescale that mainly depends on the initial surface magnetic field $B_{\rm i}$. On the other hand, the larger the initial surface magnetic field or the faster the initial spin, the smaller the precession period. And a larger deformation parameter $|\beta|$ can also cause a smaller precession period.

Based on this model, we constrain the ages of magnetars that appear (possible) periodic activities with the reasonable parameter space $(B_{\rm i},P_{\rm i})$. For FRB 180916B with 16.35-day activity period, its source age is constrained from 100 to 2000 yr. The source age of FRB 121102 with 161-day activity period is constrained from 400 to 7000 yr. Some Galactic magnetars also appear possible periodic activities in the activities of SGRs. Combining the observation of $P = 3.245 ~{\rm s}$ and $\dot{P} = 1.43\times 10^{-11}{\rm s ~s^{-1}}$ of SGR 1935+2154, we can obtain the age, initial magnetic field, and $|\beta|$ of this magnetar for a given initial period. For $P_{\rm i}=5~{\rm ms}$, one has $T_{\rm age}=2136.6~{\rm yr}$, $B_{\rm i}=5.66\times10^{14}~{\rm G}$, and $|\beta|=0.0964$. For $P_{\rm i}=10~{\rm ms}$, one has $T_{\rm age}=2103~{\rm yr}$, $B_{\rm i}=5.37\times10^{14}~{\rm G}$, and $|\beta|=0.105$. For another magnetar SGR 1806-20, although no FRBs were detected from this source, it still had a possible period of 398.2 days in X-ray band. Combining the observation of $P = 7.55 ~{\rm s}$ and $\dot{P} = 49.5\times 10^{-11}{\rm s~s^{-1}}$, we can obtain the age, initial magnetic field, and $|\beta|$ of this magnetar.
For $P_{\rm i}=5~{\rm ms}$, one has $T_{\rm age}=172.6~{\rm yr}$, $B_{\rm i}=4.40\times10^{15}~{\rm G}$, and $|\beta|=0.000865$. For $P_{\rm i}=10~{\rm ms}$, one has $T_{\rm age}=169.9~{\rm yr}$, $B_{\rm i}=4.33\times10^{15}~{\rm G}$, and $|\beta|=0.000859$. The above constraints on the source ages suggest that FRB repeaters and SGRs might be produced by young magnetars with significant deformation. 

At last, we discussed the observable probability and the active window of FRB repeaters. We notice that the radiation beam is required to be derived from the deformation axis in order to produce the observable precession signal. In this picture, we calculated the evolution of the observable probability $\eta$. We found that for a given magnetar the observable probability $\eta$ increases with the magnetar age $T_{\rm age}$ in the early stage and decreases with the magnetar age $T_{\rm age}$ in the later stage. Meanwhile, the stronger the initial magnetic field (or the faster the initial spin) the faster the evolution of the observable probability. Thus, the observed burst rate of an old source are predicted to be much lower than that of a young source, implying that FRB repeaters with high burst rates originate from young neutron stars. On the other hand, since the radiation beam is from the deformation axis, there are one or two active windows in one precession period and the distribution of the active window depends on the light of sight. Such a prediction could be tested by the future observation.

\section*{Acknowledgements}
We thank the anonymous referee for providing helpful comments and suggestions that have allowed us to improve this manuscript significantly.
We also thank Zi-Wei Li, Dongzi Li, Fa-Yin Wang, Jin-Jun Geng, and Xiang-Hua Li for their helpful discussions. This work is supported by the National Natural Science Foundation of China grant No. 12003028 and the National SKA Program of China (2022SKA0130100). The author also acknowledges supports from the ``Science \& Technology Champion Project'' (202005AB160002) and from two ``Team Projects''--the ``Innovation Team'' (202105AE160021) and the ``Top Team'' (202305AT350002), all funded by the ``Yunnan Revitalization Talent Support Program''.

\section*{Data Availability}
This theoretical study did not generate any new data. The code performed for the calculations is available upon request.

\bibliographystyle{mnras}
\bibliography{ms}

\bsp	
\label{lastpage}

\end{document}